\documentclass{article}
\usepackage[margin=3cm]{geometry}

\usepackage[colorlinks,linkcolor=black,urlcolor=black]{hyperref}
\usepackage{amsmath,amssymb,amsthm,amsfonts}

\usepackage{bbm}
\usepackage{accents}
\usepackage[mathscr]{euscript}

\numberwithin{equation}{section}

\newcommand{\id}{{\boldsymbol{\mathbbm{1}}}}
\newcommand{\cGamma}{\accentset{\circ}{\Gamma}}

\DeclareMathOperator{\tr}{tr}

\DeclareMathOperator{\Curl}{Curl}

\DeclareMathOperator{\Grad}{Grad}

\DeclareMathOperator{\Cof}{Cof}

\newcommand{\B}{\mathbf}

\begin{document}

\title{Compatibility conditions of continua using Riemann-Cartan geometry}

\author{\renewcommand{\thefootnote}{\arabic{footnote}}  
  Christian G. B\"ohmer\footnotemark[1] \ and
  Yongjo Lee\footnotemark[2]} 
\date{\today}

\footnotetext[1]{Christian G. B\"ohmer, Department of Mathematics, University College London, Gower Street, London, WC1E 6BT, UK, email: c.boehmer@ucl.ac.uk}

\footnotetext[2]{Yongjo Lee, Department of Mathematics, University College London, Gower Street, London, WC1E 6BT, UK, email: yongjo.lee.16@ucl.ac.uk}

\maketitle

\begin{abstract}
  The compatibility conditions for generalised continua are studied in the framework of differential geometry, in particular Riemann-Cartan geometry. We show that Vall\'{e}e's compatibility condition in linear elasticity theory is equivalent to the vanishing of the three dimensional Einstein tensor. Moreover, we show that the compatibility condition satisfied by Nye's tensor also arises from the three dimensional Einstein tensor which appears to play a pivotal role in continuum mechanics not mentioned before. We discuss further compatibility conditions which can be obtained using our geometrical approach and apply it to the micro-continuum theories.
\end{abstract}

\mbox{}


\mbox{}


\mbox{}

\section{Introduction}

Compatibility conditions in continuum mechanics form a set of partial differential equations which are not completely independent of each other. They may impose certain conditions among the unknown functions which are often derived by applying higher-order mixed partial derivatives to the given system of equations. They are closely related to integrability conditions.

In 1992 Vall\'{e}e~\cite{CV1992} showed that the standard Saint-Venant compatibility condition of linear elasticity, known since the mid-19th century, can be written in the convenient form 
\begin{equation}
  \label{1.1}
  \Curl\Lambda+\Cof\Lambda=0\;,
\end{equation}
where $\Lambda$ is the $3\times 3$ matrix given by 
\begin{equation}
  \label{1.2}
  \Lambda=\frac{1}{\det U}\left[U(\Curl U)^T U-\frac{1}{2}\tr\left[(\Curl U)^TU\right]U\right]\;.
\end{equation}
This formulation was based on Riemannian geometry where the metric tensor was written as $g_{\mu\nu}=U^a_\mu U^b_\nu\delta_{ab}$. Here $U$ is the right stretch tensor of the polar decomposition of the deformation gradient tensor $F=R\,U$ and $R$ is an orthogonal matrix which is the polar part. The quantities $\Curl$ and $\Cof$ in~(\ref{1.1}) are defined by
\begin{equation}\label{1.3}
  (\Curl U)_{ij} = \epsilon_{jmn}\partial_mU_{in}\;,
  \qquad\text{and}\qquad
  (\Cof U)_{ij}=\frac{1}{2}\epsilon_{ims}\epsilon_{jnt}U_{mn}U_{st}\;,
\end{equation}
and $\epsilon_{ijk}$ is the totally skew-symmetric Levi-Civita symbol.

Condition~(\ref{1.1}) was derived by finding the integrability condition of the system for the right Cauchy-Green deformation tensor $C$ which is defined by
\begin{equation}\label{1.4}
  C=(\nabla\Theta)^T(\nabla\Theta)\;.
\end{equation}
The deformation of the continuum is expressed by a diffeomorphism $\Theta:\mathcal{M}\to\mathbb{R}^3$ such that $x=X+u$ with $u$ being the displacement vector. Hence, the tensor $C$ assumes the role of a metric tensor in the given smooth manifold $\mathcal{M}$. Later~\cite{PC2007-1}, the existence of such an immersion $\Theta$ is proved that maps an open subset of $\mathbb{R}^3$ into $\mathbb{R}^3$ in which the metric tensor field defined by $C$ resides, given by $U$ in the polar decomposition $\nabla\Theta=RU$. Equation~(\ref{1.1}) was shown to be equivalent to the vanishing of the Riemann curvature tensor in this setting.

Much earlier, in 1953 Nye~\cite{JN1953} showed that there exists a curvature related rank-two tensor $\Gamma$ of the form 
\begin{equation}\label{1.5}
  \Gamma=\frac{1}{2}\tr\left(R^T\Curl R\right)\id-(R^T\Curl R)^T\;,
\end{equation}
satisfying the compatibility condition
\begin{equation}\label{1.6}
  \Curl\Gamma+\Cof\Gamma=0\;.
\end{equation}
The object $\Gamma$ is often called Nye's tensor and is written in terms of dislocation density tensor $K=R^T\Curl R$ which only depends on the orthogonal matrix $R$.

In this paper, we would like to show that these two compatibility conditions, seemingly arising from different and incomparable settings, are in fact special cases of a much broader compatibility condition which can be formulated in Riemann-Cartan geometry.

Riemann-Cartan geometry provides a suitable background when one brings the concepts of curvature and torsion to the given manifold, using the method of differential geometry in describing the intrinsic nature of defects and its classifications. Pioneering works using this mathematical framework were explored in~\cite{AC1949, EK1959, WN1958, KK1964, BB1968}. And many attempts to understand the theory of defects within the framework of the Einstein-Cartan theory were made in~\cite{HK1987, MK1992, FH2007, AY2012, AY2013}. Curvature and torsion can be regarded as the sources for disclination and dislocation densities in the theory of defects respectively. The rotational symmetries are broken by the disclination and the translational symmetries are broken by the dislocation~\cite{KK1964, RD1971, MK1977, MK1989} in Bravais lattices, the approximation of crystals into a continuum. 

It is worth noting that these geometries are commonly used in Einstein-Cartan theory~\cite{FH1976, HA2004, CB2014}, teleparallel gravity~\cite{RA2013}, gauge theories of gravity~\cite{ML2010, HK2011, MB2013} and condensed matter system~\cite{HK2004, AB2017, JN2020}. Links between micro-rotations and torsion were explored in~\cite{FH2007,CB2011-1,CB2012-1,CB2013-1, IP2019}. Recent developments in incorporating elasticity theory and spin particles using the tetrad formalism can be found in~\cite{JN2018, JN2019}.

Our paper is organised as follows: In Section 2, after introducing frame bases and co-frame bases (also will be called as tangent and co-tangent bases) together with its polar decompositions, we define various quantities including a general connection, spin connection and torsion. We will see that the Riemann tensor can be expressed in various ways using above mentioned tensors. We introduce the Einstein tensor. Then we will decompose those tensors into two parts one which is torsion-free and one that contains torsion.

In Section 3, using the tools introduced, we will derive compatibility conditions in various physical settings using a universal process. Firstly, Vall\'{e}e's result is re-derived, followed by Nye's condition. We carefully explain the connection between these two compatibility conditions and the vanishing of the Einstein tensor. Furthermore, we will show that the Nye's result is also closely linked to Skyrme theory and thus to micro-continuum theories. We briefly remark on the homotopic classification of the compatibility conditions.

Section 4 derives general compatibility conditions based on our geometric approach followed by conclusions and discussions in the final section.

\section{Tools of differential geometry}

\subsection{Frame fields}

Let us begin with a three-dimensional Riemannian manifold $\mathcal{M}$ with coordinates $x$ and let us introduce a set of basis co-vectors (or 1-forms) for the co-tangent space at some point $x \in \mathcal{M}$
\begin{equation}
  \label{2.1}
  \{e^1_\mu,e^2_\mu,e^3_\mu\} =: e^a_\mu(x) \;,
\end{equation}
where the Latin indices $a,b,\ldots$ are tangent space indices and Greek letters $\mu,\nu,\ldots$ denote coordinate indices.
This basis is often called a (co-)tetrad. The frame field consists of three orthogonal vector fields given by
\begin{equation}
  \label{2.2}
  \{E^\mu_1,E^\mu_2,E^\mu_3\} =: E^\mu_b(x) \;.
\end{equation}
These are dual basis satisfying the following orthogonality relations
\begin{equation}
  \label{2.3}
  e^a_{\mu} E_a^{\nu} = \delta_{\mu}^{\nu}\;,
  \qquad\text{and}\qquad
  e^a_{\nu} E_b^{\nu} = \delta_{b}^{a}\;.
\end{equation}
Here $\delta_{\mu}^{\nu}$ and $\delta_{b}^{a}$ are the Kronecker deltas in their respective spaces. We emphasise that for a given manifold, we can find these tangent basis locally so that we can relate different sets of tangent bases in different points by simple transformations. However, it is impossible to find a single frame field which is nowhere vanishing globally, unless the manifold is parallelisable. For example, the hair ball problem illustrates that we cannot comb the hair on the 2-sphere $S^2$ embedded in three dimensions smoothly everywhere. Hence the use of the locally defined diffeomorphism as the immersion of $\Theta:\mathcal{M}\to\mathbb{R}^3$ used in~(\ref{1.4}).

In the frame of tetrad formalism the metric tensor emerges as a secondary quantity defined in terms of $e^a_\mu$. We have
\begin{equation}
  \label{2.4}
  g_{\mu\nu} = e^a_{\mu} e^b_{\nu} \delta_{ab}\;,
\end{equation}
and recall that in Riemannian geometry this metric gives rise to an inner product between two vectors
\begin{equation}
  \label{2.5}
  A \cdot B := g_{\mu\nu} A^\mu B^\nu \;,
\end{equation}
which then naturally leads to a normed vector space.

This means we can use the co-tangent basis $e^a_\mu$ to describe the deformation from the locally flat space $\delta_{ab}$ given by the metric tensor $g_{\mu\nu}$ written in the coordinate basis. As a result, the metric tensor $g_{\mu\nu}$ is obtained from the flat Euclidean metric $\delta_{ab}$ by a set of deformations, governed by $e^a_\mu$ at each point $x\in\mathcal{M}$. Since any deformation can be regarded as a combination of rotation, shear and compression, we can apply the polar decomposition to $e^a_\mu$ as follows
\begin{equation}
  \label{2.6}
  e^a_{\mu}=R^a{}_{b} U^b_{\mu} \,.
\end{equation}
Here $R^a{}_{b}$ is an orthogonal matrix (a pure tangent space object) while the field $U^b_{\mu}$ is a symmetric and positive-definite matrix. Whenever we need to distinguish the micro-deformations from the macro-deformations, we will put a bar over the corresponding tensor. And in what follows we will often regard the matrix $R^a{}_{b}$ to be associated with micro-rotations, so that $U^b_{\mu}$ in the co-tangent basis can be thought of as the first Cosserat deformation tensor~\cite{JL2015}. This means $\overline{U}=\overline{R}^T F$. Hence the co-tangent basis is associated with the deformation gradient. 

When this decomposition is applied to~(\ref{2.4}) one arrives at
\begin{equation}
  \label{2.7}
  g_{\mu\nu} = R^a{}_c R_{ad} U^c_{\mu} U^d_{\nu} = \delta_{cd} U^c_{\mu} U^d_{\nu}\;,
\end{equation}
which shows that the metric is independent of $R^a{}_{b}$ and only depends on $U^b_{\mu}$. This is a well-known result in differential geometry, namely the metric is independent of tangent space rotations. The polar decomposition for the inverse frame is
\begin{equation}
  \label{2.8}
  E^{\mu}_a=R_a{}^b U^{\mu}_b\;,
\end{equation}
so that $U^{\mu}_b$ is the inverse of $U^a_{\mu}$, both of which are symmetric.

Consequently, the co-tangent basis~(\ref{2.1}) given a specific metric tensor~(\ref{2.7}) is not uniquely determined. Any two (co-)tetrads $\tilde{e}^a_\mu$ and $e^a_\mu$ will yield the same metric provided they are related by a rotation
\begin{equation}
  \label{2.9}
  \tilde{e}^a_\mu = Q^a{}_{b} e^b_\mu\;,
  \qquad
  Q^a{}_{b} \in \mathrm{SO}(3)\;.
\end{equation}

A metric compatible covariant derivative is introduced in differential geometry through the condition $\nabla_\alpha g_{\mu\nu} = 0$. This introduces the Christoffel symbol components $\Gamma_{\mu\nu}^{\lambda}$ as the general connection. From~(\ref{2.7}), it is natural to assume that $\nabla_\alpha e^a_\nu=0$ in the frame formalism. This, in turn, will uniquely determine the spin connection coefficients $\omega_\mu{}^a{}_b$, 
\begin{equation}
  \label{2.9-1}
  0 = \nabla_\mu e^a_\nu = \partial_\mu e^a_\nu-\Gamma^\lambda_{\mu\nu}e^a_\lambda +
  \omega_{\mu}{}^a{}_b e^b_\nu
  \qquad \Rightarrow \qquad 
  \omega_{\mu}{}^{a}{}_{b} =
  e^a_{\lambda}\Gamma^\lambda_{\mu\nu} E^\nu_b + e^a_\nu \partial_\mu E^\nu_b\;.
\end{equation}
Note that the spin connection is invariant under global rotations but not under local rotations. The derivative terms will pick up additional terms, this is of course expected when working with connections. 

The covariant derivative for a general vector $V^\mu$ is defined by
\begin{equation}\label{2.11}
  \nabla_\lambda V^\mu=\partial_\nu V^\mu+\Gamma^\mu_{\lambda\nu}V^\nu\;,
\end{equation}
where $\Gamma^\mu_{\lambda\nu}$ is a general affine connection and the lower indices in this connection are not necessarily symmetric. Being equipped with the frame (and co-frame) field, we might introduce $V^a = e^a_\mu V^\mu$ (with inverse relation $V^\mu=E^\mu_aV^a$), which denotes the tangent space components of the vector. 

Naturally, the covariant derivative of $V^a$ can be described using the spin connection, in view of~(\ref{2.9-1}). This gives
\begin{equation}
  \label{2.13}
  \nabla_\mu V^a = \partial_\mu V^a + \omega_{\mu}{}^a{}_b V^b\;,
\end{equation}
and can be extended to higher-rank objects in the same way. 

For completeness, we state the inverse of~(\ref{2.9-1}), so that the general affine connection is expressed in terms of the spin connection
\begin{equation}
  \label{2.18}
  \Gamma^\lambda_{\mu\nu} = E^\lambda_a\omega_{\mu}{}^a{}_b e^b_\nu+
  E^\lambda_a\partial_\mu e^a_\nu\;.
\end{equation}
Equations~(\ref{2.9-1}) and~(\ref{2.18}) together with the (co-)frame allow us to express geometric identities in either the tangent space or the coordinate space. In general, the non-coordinate bases $E_a = E_a^{\mu}\partial_\mu$ do not commute $\left[E_a,E_b\right] := E_a E_b - E_b E_a \neq 0$ and one introduces the object of an-holonomity as follows. Let $u$ be a smooth function, then a direct and straightforward calculation gives
\begin{align}
  \label{2.18-1}
  \left[E_a,E_b\right]u = 
   E^\mu_a E^\nu_b(\partial_\nu e^c_{\mu} - \partial_\mu e^c_\nu)E_cu\;.
\end{align}
This must be valid for the arbitrary $u$, so we can write
\begin{equation}\label{2.18-2}
  \left[E_a,E_b\right]=f^c{}_{ab}E_c\;,
\end{equation}
where the $f^c{}_{ab}$ are the so-called structure constants which are given by
\begin{equation}\label{2.18-3}
  f^c{}_{ab} = E^\mu_a E^\nu_b(\partial_\nu e^c_{\mu} - \partial_\mu e^c_\nu) \,.
\end{equation}

\subsection{Torsion and curvature}

Given an affine connection, the torsion tensor is defined by
\begin{equation}
  \label{2.18-4}
  T^\lambda{}_{\mu\nu} := \Gamma^\lambda_{\mu\nu}-\Gamma^\lambda_{\nu\mu}\;,
\end{equation}
which is the skew-symmetric part of the connection.

Throughout this paper, we will use the `decomposition' of the various tensor quantities into torsion-free parts and a separate torsion part. We will use the notation  ``$\;\circ\;$" specifically indicating the torsion-free quantities or equivalently the quantities written in terms of the metric compatible connection which is generally referred to as the Christoffel symbol.

First we decompose the connection
\begin{equation}\label{2.30}
  \Gamma^\rho_{\nu\sigma} = \cGamma^\rho_{\nu\sigma}+K^\rho{}_{\nu\sigma}\;,
\end{equation}
which introduces the contortion tensor $K^\rho{}_{\nu\sigma}$. Using the definition of torsion~(\ref{2.18-4}) we immediately have
\begin{equation}
  \label{n1}
  T^\lambda{}_{\mu\nu} = K^\lambda{}_{\mu\nu} - K^\lambda{}_{\nu\mu} \;,
\end{equation}
which one can also solve for the contortion tensor. This yields
\begin{equation}
  \label{2.20}
  K^{\lambda}{}_{\mu\nu} = \frac{1}{2}\left(T^{\lambda}{}_{\mu\nu} + T_{\nu}{}^{\lambda}{}_{\mu} - T_{\mu\nu}{}^{\lambda}\right) \:,
\end{equation}
which in turn implies the skew-symmetric property $K^{\lambda}{}_{\mu\nu}=-K_{\nu\mu}{}^{\lambda}$. Using the frame fields, we can introduce those tensors with mixed components (coordinate space and tangent space indices) which will turn out to be useful for our subsequent discussion. For example, using~(\ref{2.18}) we can write the torsion tensor in the following equivalent way. Beginning with $T^a{}_{\mu\nu}=e^a_{\lambda}T^\lambda{}_{\mu\nu}$, one arrives at
\begin{equation}\label{2.19}
  T^a{}_{\mu\nu} = \partial_\mu e^a_\nu - \partial_\nu e^a_\mu +
  \omega_{\mu}{}^{a}{}_{b}e^b_{\nu} - \omega_{\nu}{}^{a}{}_{b}e^b_{\mu} \,.
\end{equation}

The Riemann curvature tensor is defined as
\begin{equation}
  \label{2.22}
  R^\rho{}_{\sigma\mu\nu} :=
  \partial_\mu\Gamma^\rho_{\nu\sigma}-\partial_\nu\Gamma^\rho_{\mu\sigma}+
  \Gamma^\rho_{\mu\lambda}\Gamma^\lambda_{\nu\sigma}-\Gamma^\rho_{\nu\lambda}\Gamma^\lambda_{\mu\sigma}\;.
\end{equation}
Using~(\ref{2.18}), we can rewrite the Riemann tensor with mixed indices
\begin{equation}
  \label{2.24}
  R^a{}_{b\mu\nu} = e^a_\rho R^\rho{}_{\sigma\mu\nu} E^\sigma_b
\end{equation}
where the Riemann tensor is now expressed in terms of the spin connections only
\begin{equation}\label{2.25}
  R^a{}_{b\mu\nu} = 
  \partial_{\mu} \omega_{\nu}{}^{a}{}_{b} - \partial_{\nu} \omega_{\mu}{}^{a}{}_b +
  \omega_{\mu}{}^{a}{}_{e} \omega_{\nu}{}^{e}{}_{b} - \omega_{\nu}{}^{a}{}_{e} \omega_{\mu}{}^{e}{}_{b}\;.
\end{equation}
In addition to the skew-symmetry in the last two indices in the Riemann tensor, this satisfies
\begin{equation}\label{2.29}
R_{ab\mu\nu}=-R_{ba\mu\nu}\;.
\end{equation}

As a consequence of~(\ref{2.30}) we apply the same concept to the spin connection to write the decomposition
\begin{equation}
  \label{2.30-1}
  \omega_{\mu}{}^a{}_b = \accentset{\circ}{\omega}_{\mu}{}^a{}_b + K^a{}_{\mu b}\;,
\end{equation}
where we used $K^a{}_{\mu b} = e^a_\nu K^\nu{}_{\mu\sigma}E^\sigma_b$. At first sight, the choice of index positions appears odd but ensures agreement with~(\ref{n1}).

Inserting the relation~(\ref{2.30}) into~(\ref{2.22}) gives rise to the decomposition of the Riemann tensor,
\begin{equation}\label{2.31}
  R^\rho{}_{\sigma\mu\nu} = \accentset{\circ}{R}^\rho{}_{\sigma\mu\nu} +
  \left[\accentset{\circ}{\nabla}_\mu K^\rho{}_{\nu\sigma} - \accentset{\circ}{\nabla}_\nu K^\rho{}_{\mu\sigma}+
    K^\rho{}_{\mu\lambda}K^\lambda{}_{\nu\sigma}-K^\rho{}_{\nu\lambda}K^\lambda{}_{\mu\sigma}\right]\;,
\end{equation}
where the Riemann tensor $\accentset{\circ}{R}^\rho{}_{\sigma\mu\nu}$ is computed using the connection $\accentset{\circ}{\Gamma}^\rho_{\mu\nu}$ entirely.

We note that, for a general vector $V^\rho$, in the coordinate basis, the covariant derivative can be rewritten using~(\ref{2.30}) such that 
\begin{align}
  \nabla_\mu V^\rho = \accentset{\circ}{\nabla}_\mu V^\rho + K^\rho{}_{\mu\nu}V^\nu\;.
  \label{2.32}
\end{align}
This relates the general covariant derivative $\nabla_\mu$ and the torsion-free, metric compatible covariant derivative $\accentset{\circ}{\nabla}_\mu$ used in~(\ref{2.31}). In addition to~(\ref{2.30}) and~(\ref{2.30-1}), we can regard the contortion tensor on the right-hand side as the connection between these two covariant derivatives.

\subsection{Einstein tensor in three-dimensional space}

We define a rank 2 quantity based on the spin connection by
\begin{equation}
  \label{2.35}
  \Omega_{c\mu} := -\frac{1}{2}\varepsilon_{abc}\omega_{\mu}{}^{ab}\;,
\end{equation}
which is equivalent to write
\begin{equation}
  \label{2.36}
  \omega_{\mu}{}^{ab}=-\varepsilon^{abc}\Omega_{c\mu}\;.
\end{equation}
We would like to note that this construction is tied to $\mathbb{R}^3$. The Levi-Civita symbol in $n$ dimensions maps the spin connection from a rank 3 object to a rank $n-1$ object, namely $\varepsilon_{abc \ldots n}\omega_{\mu}{}^{ab}$. Only in 3 dimensions one would arrive at a rank 2 object. In the following it will turn out that $\Omega_{c\mu}$ plays a crucial role in establishing our compatibility conditions. The same approach was applied to the torsion tensor in~\cite{CB2011-1} where the setting was also $\mathbb{R}^3$.

We substitute~(\ref{2.36}) into the Riemann tensor~(\ref{2.25}) and find
\begin{equation}\label{2.37}
  R^a{}_{b\mu\nu} =
  \varepsilon^{sa}{}_{b}(
  -\partial_{\mu} \Omega_{s\nu} +
  \partial_{\nu} \Omega_{s\mu}) +
  \varepsilon^{sa}{}_{e}\varepsilon^{te}{}_{b}(
   \Omega_{s\mu} \Omega_{t\nu} -
   \Omega_{s\nu} \Omega_{t\mu}) \;.
\end{equation}
Next, we define the following rank 2 tensor, constructed from the Riemann tensor
\begin{equation}
  \label{2.38}
  G^{\sigma c}=-\frac{1}{4}\varepsilon^{abc} R_{ab\mu\nu} \varepsilon^{\mu\nu\sigma}\;,
\end{equation}
where we recall that the Riemann curvature tensor is skew-symmetric in the first and second pairs of indices. Let us emphasise again that this construction is only possible in three dimensions, otherwise we would need to introduce a different rank in the Levi-Civita symbol.

Inserting~(\ref{2.37}) into~(\ref{2.38}) using the formulae $\varepsilon^{abc} \varepsilon_{sab}  = 2 \delta_s^c$ and $\varepsilon_{sae} \varepsilon^{abc} \varepsilon^{te}{}_{b} = - \varepsilon^{tc}{}_{s}$, we obtain
\begin{equation}
  \label{2.39}
  G^{\sigma c}=
  \varepsilon^{\mu\nu\sigma}\partial_{\mu} \Omega^c{}_{\nu} + \frac{1}{2}
  \varepsilon^{cst}\varepsilon^{\sigma\mu\nu} \Omega_{s\mu} \Omega_{t\nu}
\end{equation}
which can be written in the convenient form 
\begin{equation}
  \label{2.40}
  G^{\sigma c}=(\Curl\Omega)^{c\sigma}+(\Cof\Omega)^{c\sigma}\;.
\end{equation}
The quantity $G_{\sigma c}$ is in fact the Einstein tensor in three-dimensional space. This can be shown using~(\ref{2.38}) and~(\ref{2.24}) explicitly to obtain
\begin{equation}\label{2.45}
G_{\tau\lambda}=R_{\tau\lambda}-\frac{1}{2}\delta_{\tau\lambda}R\;.
\end{equation}
Here $R_{\tau\lambda}$ is the Ricci tensor defined by $R_{\tau\lambda} = R^\sigma{}_{\tau\sigma\lambda}$ and the trace of Ricci tensor is the Ricci scalar $R$. It is well known that in three dimensions, the Riemann tensor, the Ricci tensor and the Einstein tensor have the same number of independent components, namely nine, provided torsion is included. One can readily verify that
\begin{equation}\label{2.46}
  R^a{}_{b\mu\nu} = 0
  \quad\Leftrightarrow\quad
  R_{\tau\lambda} = 0
  \quad\Leftrightarrow\quad
  G_{\tau\lambda} = 0 \;.
\end{equation}
In other words, the vanishing curvature means vanishing Einstein tensor in three dimensions. Let us emphasise here that the particular representation of the Einstein tensor given in~(\ref{2.40}) will be of importance for what follows.

\section{Compatibility conditions}

\subsection{Vall\'{e}e's classical result}

We consider the torsion-free spin connection $\accentset{\circ}{\Omega}_{c\mu}=-\frac{1}{2}\varepsilon_{abc}\accentset{\circ}{\omega}_{\mu}{}^{ab}$ with the metric tensor~(\ref{2.4}). The affine connection in torsion-free spaces is conventionally expressed by the metric compatible Levi-Civita connection
\begin{equation}
  \label{3.1}
  \accentset{\circ}{\Gamma}^{\alpha}_{\beta\gamma} = \frac{1}{2}g^{\alpha\sigma}
  \left(\partial_\gamma g_{\sigma\beta}+\partial_\beta g_{\gamma\sigma}-\partial_\sigma g_{\beta\gamma}\right)\;.
\end{equation}
The torsion-free spin connection in terms of the Levi-Civita connection is simply
\begin{align*}
  \accentset{\circ}{\omega}_{\mu}{}^a{}_b &=
  e^a_{\lambda}\accentset{\circ}{\Gamma}^\lambda_{\mu\nu}E^\nu_b+e^a_\nu \partial_\mu E^\nu_b\\
  &=\frac{1}{2}e^a_\lambda g^{\lambda\tau}
  \left(\partial_\nu g_{\tau\mu}+\partial_\mu g_{\tau\nu}-\partial_\tau g_{\mu\nu}\right)E^\nu_b +
  e^a_\nu \partial_\mu E^\nu_b\;,
\end{align*}
where we used~(\ref{2.9-1}). Inserting the explicit expression for the metric tensor~(\ref{2.4}) will give, after a lengthy but simple calculation
\begin{equation}
  \label{3.2}
  \accentset{\circ}{\omega}_\mu{}^a{}_b =
  \frac{1}{2}E_b^{\sigma} \left(\partial_\sigma e_\mu^a-\partial_\mu e_\sigma^a \right) -
  \frac{1}{2}\delta^{ad} \delta_{fb} E_d^\sigma
  \left(\partial_\sigma e_\mu^f - \partial_\mu e_\sigma^f \right) +
  \frac{1}{2}\delta^{ad} g_{\mu\sigma}
  \left(\partial_d E_b^\sigma - \partial_b E_d^\sigma \right)\;.
\end{equation}
Here we used the notation $\partial_a = E_a^\sigma \partial_\sigma$. Furthermore, we can write the spin connection in terms of polar decomposition of co-frame field basis $e^a_\mu=R^a{}_{b} U^b_\mu$ to write $\accentset{\circ}{\omega}_\mu{}^{ab}$ entirely in terms of $R^a{}_{b}$ and $U^b_\mu$ and its derivatives. The resulting expression will be further simplified if we consider the cases $R^a{}_{b}=\delta^a_b$ and $U^d_\nu=\delta^d_\nu$ separately, to see whether these will lead to the desired compatibility conditions.

First, when $R^a{}_{b}=\delta^a_b$ after multiplying $\epsilon_{abc}$ to both sides of~(\ref{3.2}), we have
\begin{equation}\label{3.3}
\epsilon_{abc}\accentset{\circ}{\omega}_\mu{}^{ab}=\epsilon_{abc}\epsilon_{\sigma\mu\nu}U^{a\nu}(\Curl U)^{b\sigma}-\frac{1}{2}\epsilon_{abc}\epsilon_{\sigma\tau\rho}U^{a\rho}U^{b\sigma}(\Curl U)_f{}^{\tau}U^f_\mu\;.
\end{equation}
We can extract the determinant of $U$ from the first and the second term in the right-hand side of this,
\begin{equation}\label{3.4}
\begin{split}
  \epsilon_{abc}\epsilon_{\sigma\mu\nu}U^{a\nu}(\Curl U)^{b\sigma}&=\frac{6}{\det U}\;\left[U(\Curl U)^TU\right]_{c\mu}\\
  \epsilon_{abc}\epsilon_{\sigma\tau\rho}U^{a\rho}U^{b\sigma}(\Curl U)_f{}^{\tau}U^f_\mu&=
  \frac{6}{\det U}\;U_{c\mu}\tr\left[(\Curl U)^TU\right]\;.
\end{split}
\end{equation}
Therefore, we find
\begin{equation}
  \label{3.5}
  \accentset{\circ}{\Omega}_{c\mu}=-3\cdot\frac{1}{\det U}\left[U(\Curl U)^T U-\frac{1}{2}\tr\left[(\Curl U)^TU\right]U\right]_{c\mu}\;.
\end{equation}

The vanishing Riemann tensor in three-dimensional space ensures the vanishing Ricci tensor, hence the vanishing of the Einstein tensor $\accentset{\circ}{G}_{\mu c}=0$ as stated in~(\ref{2.46}). This leads to the compatibility condition in the torsion-free space of vanishing Riemann curvature, with the help of~(\ref{2.40}),
\begin{equation}\label{3.6}
  \Curl\accentset{\circ}{\Omega}+\Cof\accentset{\circ}{\Omega}=0\;.
\end{equation}
We can rescale $-\frac{1}{3}\accentset{\circ}{\Omega}=\Lambda_U$ to match the Vall\'{e}e's result~\cite{CV1992} exactly
\begin{equation}\label{3.7}
\Lambda_U=\frac{1}{\det U}\left[U(\Curl U)^T U-\frac{1}{2}\tr\left[(\Curl U)^TU\right]U\right]\;,
\end{equation}
which reads
\begin{equation}\label{3.8}
\Curl\Lambda_U+\Cof\Lambda_U=0\;.
\end{equation}

The elastic deformation is nothing but the diffeomorphism described by a metric tensor with associated metric compatible connection $\accentset{\circ}{\Gamma}^{\alpha}_{\beta\gamma}$ as the fundamental measure of the deformation. Then, the prescription of elastic deformations requires vanishing curvature and torsion, hence the compatibility conditions~(\ref{3.6}).

We should also note the results of Edelen~\cite{DE1990} where compatibility conditions were derived using Poincar\'{e}'s lemma. This resulted in the vanishing Riemann curvature 2-form, equation~(3.3) in~\cite{DE1990} while assuming a metric compatible connection, equation~(3.4) in~\cite{DE1990}. These conditions explicitly contained torsion due to the affine connection being non-trivial but curvature free. 

\subsection{Nye's tensor and its compatibility condition}

In the following we set $U^c_\mu=\delta^c_\mu$ but assume a non-trivial rotation matrix $R^a{}_{b}$, this is the opposite to the previous case. The compatibility condition from~(\ref{3.2}) becomes
\begin{equation}
  \label{3.9}
  \Curl\Lambda_R+\Cof\Lambda_R=0\;,
\end{equation}
where the quantity $\Lambda_R$ is given by
\begin{equation}
  \label{3.10}
  \Lambda_R= R(\Curl R)^T R-\frac{1}{2}\tr\left[(\Curl R)^TR\right]R \;.
\end{equation}
This is formally identical to replacing $U^c_\mu$ with $R^a{}_{b}$ in~(\ref{3.7}) and using $\det R^a{}_{b} =+1$.

It turns out that the quantity $\Lambda_R$ is (up to a minus sign) Nye's tensor $\Gamma$ which is known to satisfy the compatibility condition~(\ref{3.9}). This is quite a remarkable result which follows immediately from our geometrical approach to the problem. 

We emphasise that the metric tensor is independent of the rotations which implies that $U^c_\mu=\delta^c_\mu$ yields a vanishing (torsion-free) Levi-Civita connection $\accentset{\circ}{\Gamma}^\alpha_{\beta\gamma}$. Consequently the Levi-Civita part of the curvature tensor vanishes identically, $\accentset{\circ}{R}^\rho{}_{\sigma\mu\nu} = 0$. Nonetheless, the non-trivial rotational part of the frame contributes to the curvature tensor $R^\rho{}_{\sigma\mu\nu}$ in~(\ref{2.31}) through the contortion tensor, since the general connection $\Gamma^\rho_{\mu\nu}$ does not vanish in this case. The compatibility condition simply ensures that the micropolar deformations do not induce curvature into the deformed body. Most importantly, torsion is not assumed to vanish and the rotation matrices $R^a{}_{b}$ become dynamic and non-trivial.

Let us note that in the space where $\accentset{\circ}{\Gamma}^\lambda_{\mu\nu}=0$, or equivalently $U^c_\mu=\delta^c_\mu$ and non-vanishing torsion, the general connection becomes the contortion. Moreover, by setting $\omega_\mu{}^a{}_b = 0$ in~(\ref{2.18}) this yields
\begin{align}\label{3.11}
  \Gamma^\lambda_{\mu\nu} = (R_a{}^b \delta_b^\lambda) \partial_\mu (R^a{}_c \delta^c_\nu) =
  \delta_b^\lambda \delta^c_\nu (R_a{}^b \partial_\mu R^a{}_c) =
  \delta_b^\lambda \delta^c_\nu \delta^d_\mu (R_a{}^b \partial_d R^a{}_c) \;.
\end{align}
The final term in the brackets is recognised to be the second Cosserat tensor when written in index free notation $R^T \Grad R$, see for instance~\cite{JL2015}. This tensor is sometimes denoted by $K$, in order to avoid confusion with our contortion tensor we shall refrain from using this notation.

In the following we will briefly discuss how the compatibility condition for Nye's tensor can also be derived directly without referring to the general result~(\ref{2.40}). In order to have completely vanishing curvature tensor~(\ref{2.31}) with $U^c_\mu=\delta^c_\mu$ we note:
\begin{enumerate}
\item The Levi-Civita connection $\accentset{\circ}{\Gamma}^\rho_{\mu\nu} = 0$ and $\accentset{\circ}{R}^\rho{}_{\sigma\mu\nu} = 0$ in~(\ref{2.31}).
\item The connection and contortion tensors becomes identical using~(\ref{2.30}), as in the case of~(\ref{3.11}).
\item We can replace $\accentset{\circ}{\nabla}_\mu$ with $\partial_\mu$ in~(\ref{2.31}).
\end{enumerate}
Under these circumstances, the Riemann tensor~(\ref{2.31}) reduces to
  \begin{equation}
    \label{3.15}
    R^\rho{}_{\sigma\mu\nu}=
    \partial_\mu K^\rho{}_{\nu\sigma} - \partial_\nu K^\rho{}_{\mu\sigma} +
    K^\rho{}_{\mu\lambda}K^\lambda{}_{\nu\sigma}-K^\rho{}_{\nu\lambda}K^\lambda{}_{\mu\sigma}\;.
\end{equation}

We introduce, similar to~(\ref{2.35}), the dislocation density tensor
\begin{equation}
  \label{3.16}
  K_{\lambda\sigma} := \epsilon_\sigma{}^{\mu\nu}K_{\lambda\mu\nu}\;,
\end{equation}
which for our explicit choice of contortion in~(\ref{3.11}) we can write
\begin{equation}
  \label{3.18}
  K_{\lambda\sigma} = \epsilon_\sigma{}^{\mu\nu} \delta_{\lambda b} R_a{}^{b}
  \partial_\mu R^a{}_{c} \delta^c_\nu =
  \left(R^T\Curl R\right)_{\lambda\sigma}\;.
\end{equation}
For Nye's tensor, we contract the first and third index of the contortion tensor
\begin{equation}
  \label{3.19}
  \Gamma_{\lambda\nu} := -\frac{1}{2}
  \epsilon_\lambda{}^{\rho\sigma}K_{\rho\nu\sigma}\;.
\end{equation}
In turn, the relation between Nye's tensor and contortion becomes $\Gamma_{\lambda\nu}\epsilon^\lambda{}_{\alpha\beta}=-K_{\alpha\nu\beta}$. From this, the contortions can be substituted into~(\ref{3.15}) to write the Riemann curvature in terms of Nye's tensor. This immediately yields
\begin{align}
  \label{3.24}
  \epsilon^\delta{}_{\rho\sigma} \partial_\rho\Gamma_{\alpha\sigma} +
  \frac{1}{2}\epsilon^{\tau\eta}{}_{\alpha} \epsilon^\delta{}_{\rho\sigma}
  \Gamma_{\tau\rho}\Gamma_{\eta\sigma} &= 0\;, \\
  \Leftrightarrow \quad
  \left(\Curl\Gamma\right)_{\alpha\delta}+\left(\Cof\Gamma\right)_{\alpha\delta} &= 0\;.
  \label{3.25}
\end{align}
This is our second compatibility condition written in terms of Nye's tensor, for the vanishing curvature and nonzero torsion space.

We note that combining~(\ref{3.18}) and~(\ref{3.19}) together leads to the usual expression of Nye's tensor
\begin{equation}
  \label{3.21}
  \Gamma_{\lambda\nu}=\frac{1}{2}\tr\left(R^T\Curl R\right)\delta_{\lambda\nu}-(R^T\Curl R)^T{}_{\lambda\nu}\;.
\end{equation}

\subsection{Skyrme's theory with compatibility condition}

In a series of papers~\cite{TS1958,TS1959,TS1961-1} Skyrme introduced a nonlinear field theory for describing strongly interacting particles. This work has motivated many subsequent studies and noted some interesting links between baryon numbers, the sum of the proton and neutron numbers, and topological invariants in field theory. Following Skyrme's notation, the key variable is the field
\begin{equation}
  B^\alpha_\mu = -\frac{1}{4}\epsilon_{\alpha\beta\gamma}g^{\beta\delta}
  \frac{\partial}{\partial x_\mu} g_{\gamma\delta} \:,
\end{equation}
where $g$ denotes an orthogonal matrix. Now, using $R^a{}_b$ to denote the orthogonal matrix instead, we note that the field $B^\alpha_\mu$ is related to $R^T \partial_\mu R$ which is generally referred to as the second Cosserat tensor~\cite{JL2015} so that we immediately note a close similarity between Skyrme's nonlinear field theory and Cosserat elasticity. It was noted in~\cite{TS1961-1} that the `covariant curl of' $B$ vanishes identically
\begin{equation}
  \partial_\nu B^\alpha_\mu - \partial_\mu B^\alpha_\nu - 2\epsilon_{\alpha\beta\gamma}
  B^\beta_\mu B^\gamma_\nu = 0 \;.
  \label{covcurl}
\end{equation}
If we now contract this equations with $\epsilon^{\mu\nu\sigma}$ we will recognise the final product as the cofactor matrix of $B$ while the first becomes the matrix Curl. Therefore, the `covariant curl' equation~(\ref{covcurl}) is equivalent to
\begin{equation}
  \Curl B + \Cof B = 0 \;.
  \label{covcurl2}
\end{equation}
Perhaps unsurprisingly, at this point, a direct calculation shows that Skyrme's field is in fact Nye's tensor. Using our notation, we have 
\begin{equation}
  \label{3.27}
  B_{aj} = -\frac{1}{2}\epsilon_{ai}{}^{s}\Gamma^i_{js} = -
  \frac{1}{2}\epsilon_{ai}{}^{s}\left(
  \accentset{\circ}{\Gamma}^i{}_{js}+K^i{}_{js}
  \right) = -
  \frac{1}{2}\epsilon_{ai}{}^{s} K^i{}_{js}=\Gamma_{aj}\;.
\end{equation}
In the third step, we used the condition $U^c_\mu=\delta^c_\mu$, hence $\accentset{\circ}{\Gamma}^\rho_{\mu\nu}=0$. As in the previous subsection, we can derive this equation explicitly by requiring the complete Riemann curvature tensor~(\ref{2.22}) to vanish. Together the assumption of a trivial metric tensor with non-trivial frame field, this is equivalent to satisfying~(\ref{3.15}). Consequently, Skyrme's condition~(\ref{covcurl}) is in fact~(\ref{3.9}) or equivalently~(\ref{3.25}).

Since Skyrme's variable is in fact Nye's tensor in three dimensions, it becomes clear that it also must have a relation to a topological invariant. In the context of Cosserat elasticity this connection has been noted in~\cite{HT1981, HT1982} where it is shown that the winding number can be written as the integration of the determinant of the Nye's tensor over all space defined in the given manifold $\mathcal{M}$ to write
\begin{equation}
  \label{3.27-1}
  n = -\frac{1}{(4\pi)^2} \int_{\mathcal{M}} \det\Gamma\, d^3x\;,\qquad n\in\mathbb{Z}\;.
\end{equation}
The factor of $2\pi^2$ is due to the surface area of $S^3$. This can be understood by recalling that a unit vector $\hat{v} \in \mathbb{R}^4$ has 3 independent components, hence $\hat{v} \in S^3$, which in turn allows one to define orthogonal matrices through $\hat{v}$. The determinant of the Nye tensor is simply related to the determinant of the induced metric of $S^3$ and thus relates to its volume. Notably, in~\cite{CB2012-1} the form of integration using contortion one-forms gives
\begin{equation}
  \label{3.27-2}
  n=\frac{1}{96\pi^2}\int_{\mathcal{M}}\tr\left(K\wedge K\wedge K\right)\;,\qquad n\in\mathbb{Z}\;,
\end{equation}
which can be derived from a Chern-Simons type action in terms of contortion, seen as gauge fields,
\begin{equation}
  \label{3.27-3}
  S=\frac{1}{4\pi}\int_{\mathcal{M}}\tr(K\wedge dK+\frac{2}{3}K\wedge K\wedge K)\;.
\end{equation}
The two integrations~(\ref{3.27-1}) and~(\ref{3.27-2}) can be shown to be identical using the relation~(\ref{3.19}). The agreement of the compatibility conditions for Skyrme's field and Nye's tensor is, by no means, accidental. In particular, by varying the action~(\ref{3.27-3}) with respect to contortion, one arrives at the equation of motion
\begin{equation}\label{3.27-3a}
dK+K\wedge K=0\;,
\end{equation}
which agrees with~(\ref{3.15}), the vanishing Riemann tensor with nonzero torsion, see again~\cite{DE1990}.

One might get the impression from~(\ref{3.15}) that non-vanishing curvature is induced by the non-vanishing contortion or  torsion. However, this is not the case. As indicated in~(\ref{3.27-3a}), contortion is of Maurer-Cartan form $K=R^{T}dR$ which satisfies the Maurer-Cartan equation $dK=-K\wedge K$. In our setting we considered two kinds of compatibility conditions so far, namely we have
\begin{alignat}{5}
  \label{3.27-4}
  U^c_\mu &= \delta^c_\mu
  &\quad &\Rightarrow &\quad
  \accentset{\circ}{\Gamma}^\lambda_{\mu\nu} &=\accentset{\circ}{\Omega}_{c\mu}=
  \accentset{\circ}{\omega}_{\mu}{}^a{}_b = 0
  &\quad &\Rightarrow &\quad
  \accentset{\circ}{R}^\rho{}_{\sigma\mu\nu} &= 0 \text{ and } R^\rho{}_{\sigma\mu\nu}=0 \;,\\
  R^a{}_{b} &= \delta^a_b
  &\quad &\Rightarrow &\quad
  K_{\lambda\nu} &= \Gamma_{\lambda\nu} = K_{\alpha\nu\beta} = 0
  &\quad &\Rightarrow &\quad
  T^\lambda{}_{\mu\nu} &= 0\;.
\end{alignat}
The converse is not true in general as will be shown in Section~\ref{sec:geo} when deriving the general form of the compatibility conditions.

Finally we note that there has also been some mathematical interests in this topic, see for instance~\cite{ME1986, ME1992} where Skyrme's model was studied using a variational approach. The key challenge was to find minimisers subject to appropriate boundary conditions which yield soliton solutions. Discrete topological sectors according to these solutions will lead to the topological number in accordance with the distinct homotopy classifications. These topological invariants can be found in diverse physical systems with order parameters describing the `defects' of distinct nature such as monopoles, vortices and domain walls~\cite{AU2001, AR2011}. Certain `optimal' properties of orthogonal matrices in the context of Cosserat elasticity were studied in~\cite{PN2014-2,AF2017-3,PN2019,LB2019}

\subsection{Eringen's compatibility conditions}

Generalised continua are characterised by replacing the idealised material point with an object with additional micro-structure. The inner structure is described by directors which can undergo deformations such as rotation, shear and compression which introduces 9 additional degrees of freedom. The first ideas along those lines go back to the Cosserat brothers who, in 1909, first considered such theories~\cite{EC}. A comprehensive account of micro-continuum theories can be found in~\cite{AE1}. In particular, micropolar theory describes the rigid micro-rotation for the micro-element deformation. Nonlinear problems in generalised continua were studied rigorously for instance in~\cite{PN2006, PN2008, PN2013, PN2015-1, PN2015-2, PN2016}.

Let us begin by briefly recalling the basic notation used in~\cite{AE1}. First, we introduce strain measures
\begin{alignat}{2}
  \mathfrak{C}_{KL}&=\frac{\partial x_k}{\partial X_K}\mathfrak{X}_{Lk}\;,
  &\qquad \mathscr{C}_{KL}&=\chi_{kK}\chi_{kL}=\mathscr{C}_{LK}\;,\\
  \Gamma_{KLM}&=\mathfrak{X}_{Kk}\frac{\partial\chi_{kL}}{\partial X_M}\;,
  &\qquad \Gamma_{KL}&=\frac{1}{2}\epsilon_{KMN}\Gamma_{NML}\;.
\end{alignat}
The tensors $\chi_{kK}=\partial\xi_k/\partial\Xi_K$ and $\mathfrak{X}_{Kk}=\partial\Xi_K/\partial\xi_k$ are called micro-deformation tensors and inverse micro-deformation tensors with the directors $\Xi_K$ and $\xi_k$ in material coordinate $X_K$ and spatial coordinate $x_k$ respectively. These satisfy orthogonal relations $\chi_{kK}\mathfrak{X}_{Kl}=\delta_{kl}$ and $\mathfrak{X}_{Kl}\chi_{lL}=\delta_{KL}$.

Now, these micro-deformation tensors can be decomposed into rotation and stretch parts, again the polar decomposition, as we did in bases $e^a_\mu$ and $E^\nu_a$. For example, after changing indices in accordance with our convention, we can rewrite
  \begin{align}
  \chi^a{}_c&=\bar{R}^a{}_b\bar{U}^b_c\;,\qquad\mathfrak{X}_a{}^c=\bar{R}_a{}^b\bar{U}_b^c\;,
  \label{3.29a}\\
  \mathfrak{C}^\mu_a&=\mathfrak{X}_a{}^cF_c{}^\mu=\bar{R}_a{}^b\bar{U}_b^cR_c{}^dU_d^\mu\;,
  \label{3.29b}\\
  \mathscr{C}_{bc}&=\chi^a{}_b\chi_{ac}=\bar{R}^a{}_c\bar{U}^c_b\bar{R}_{ad}\bar{U}^d_c\;,
  \label{3.29c}\\
  \Gamma_{klm}&=\chi^a{}_k\partial_m\chi_{al}=\bar{R}^a{}_b\bar{U}^b_k\partial_m(\bar{R}_{ac}\bar{U}^c_l)\;,
  \label{3.29d}
\end{align}
in which we used bars over the the micro deformations and used definition for the (macro)deformation gradient tensor $F$ with its polar decomposition into macro-rotation and macro-stretch.

The compatibility conditions for the micromorphic body~\cite{AE1} are given by
\begin{align}
  \epsilon_{KPQ}\left(\partial_Q\mathfrak{C}_{PL}+\mathfrak{C}_{PR}\Gamma_{LRQ}\right)&=0
  \label{3.30a}\\
  \epsilon_{KPQ}\left(\partial_Q\Gamma_{LMP}+\Gamma_{LRQ}\Gamma_{RMP}\right)&=0
  \label{3.30b}\\
  \partial_M\mathscr{C}_{KL}-\left(\Gamma_{PKM}\mathscr{C}_{LP}+\Gamma_{PLM}\mathscr{C}_{KP}\right)&=0
  \label{3.30c}
\end{align}
where $\partial_M=\partial/\partial X_M$. It is evident from~(\ref{3.29d}) that the wryness tensor $\Gamma_{KLM}$ can be viewed as  the contortion tensors in differential geometry, so we can make a replacement $\Gamma_{PKM}\to K^P{}_{MK}$, hence the compatibility condition~(\ref{3.30c}) now becomes
\begin{equation}
  \label{3.31}
  \partial_M\mathscr{C}_{KL} - K^P{}_{MK} \mathscr{C}_{PL} - K^P{}_{ML} \mathscr{C}_{KP} = 0\;.
\end{equation}
Using the decomposition~(\ref{2.30}) with $\accentset{\circ}{\Gamma}^P_{MK}=0$, this will further reduce to
\begin{equation}
  \label{3.33}
  \nabla_M\mathscr{C}_{KL}=0\;.
\end{equation}
This condition is now equivalent to assuming a metric compatible covariant derivative, see after~(\ref{2.9}), one of our central assumption of the geometrical approach. 

Next, we consider condition~(\ref{3.30b}). We have
\begin{equation}\label{3.34}
  \partial_QK^L{}_{PM} - \partial_PK^L{}_{QM} + K^L{}_{QR}K^R{}_{PM} - K^L{}_{PR}K^R{}_{QM} = 0\;.
\end{equation}
The left-hand side of this is in the form of the Riemann curvature tensor~(\ref{3.15}), hence this condition is equivalent to $R^L{}_{MQP} = 0$. This is our second geometrical condition that led to the compatibility conditions.

Lastly, for~(\ref{3.30a}) one writes
\begin{equation}\label{3.36}
  \epsilon_{KPQ}\left(\partial_Q\mathfrak{C}_{PL}+K^L{}_{QR}\mathfrak{C}_{PR}\right)=0\;,
\end{equation}
which is known as the compatibility condition for the disclination density tensor. After some algebraic manipulation this final condition can be written as
\begin{equation}\label{3.36a}
  \nabla_Q \mathfrak{C}^L_P - \nabla_P \mathfrak{C}^L_Q + T^R{}_{PQ} \mathfrak{C}^L_R = 0 \;,
\end{equation}
and can be seen as the defining equation for torsion on the manifold.

The above shows that the setting of Riemann-Cartan geometry appears to be very well suited to study a micromorphic continuum.

\subsection{Homotopy for the compatibility condition}

In~\cite{PC2005}, it is shown that the existence of the metric tensor field~(\ref{1.4}) for a given immersion $\Theta:\Omega\to\mathbb{E}^3$ requires the condition $R^\rho{}_{\sigma\mu\nu}=0$ in $\Omega\subset\mathbb{R}^3$ and $\Omega$ to be simply-connected. It is further shown to be necessary and sufficient.

If the subset of the given manifold is just connected subset, then $\Theta$ is unique up to isometry of Euclidean space $\mathbb{E}^3$ to ensure the existence of the metric field
\begin{align}
  C=(\nabla\tilde{\Theta})^T(\nabla\tilde{\Theta})\;,
\end{align}
where $\Theta=Q\tilde{\Theta}+T$ for $Q\in\mathrm{SO(3)}$ and $T$ is translation.

Now, we might wish to establish how many compatibility conditions, or more precisely, how many classifications of such compatibility conditions are derivable from the condition $R^\rho{}_{\sigma\mu\nu}=0$? One possible approach to answer this question would be the consideration of the homotopy classification $\pi_n(M)$, where $n$ is the dimension of the $n$-sphere $S^n$, the probe of the defects in the space $M$ in which the order parameter is defined. In our case, we can put the order parameter to simply be the tetrad field $e^a_\mu$ so that $M=\mathrm{SO(3)}$.

It is well-known that the dislocation or equivalently the torsion can be measured by following a small closed path in the  crystal lattice structure, and the curvature can be computed in a similar manner. We can put $n=1$ to consider the fundamental group for $\mathrm{SO(3)}$, which is a homotopy group for the line defects in three dimensions
\begin{align}
  \pi_1(\mathrm{SO(3)})\cong\mathbb{Z}_2\;.
\end{align}
This suggests that we can have two distinct classifications for the compatibility conditions under $R^\rho{}_{\sigma\mu\nu}=0$. One of them is to the trivial class, the elastic regime so that all elastic deformations belong to the same compatible condition. And the non-trivial classification is for the micro-structure description where one is only dealing with micro-deformations. Similar analysis can be found in~\cite{DF1966, MK1977, RS1977}.

Interestingly, in some simplified Skyrme models~\cite{MN1987}, the homotopy class $\pi_4(\mathrm{SO(3)})$ is identified with $\pi_1(\mathrm{SO}(3))$. Since $\mathrm{SO}(3)$ is not simply connected, it is straightforward to see that its fundamental group is isomorphic to $\mathbb{Z}_2$. Further, using $J$-homomorphism, we can state
\begin{align}
  \pi_4(\mathrm{SO(3)})\cong\pi_1(\mathrm{SO(3)})\cong\mathbb{Z}_2\;.
\end{align}
This characterises the equivalent classes of the compatibility conditions, hence the possible solutions for the system in describing the deformations, as below:
\begin{align*}
\{0\}:\;&\text{Configurations that can be continuously deformed uniformly via diffeomorphism.}\\
\{1\}:\;&\text{Configurations that cannot be continuously deformed in a way of $\{0\}$.}
\end{align*}
The elastic compatibility condition including the Vall\'{e}e's result~(\ref{1.1}) falls into the classification $\{0\}$: vanishing curvature and torsion.  The conditions by Nye~(\ref{1.6}), Skyrme field~(\ref{covcurl2}) and micropolar case~(\ref{3.30b}) belong to $\{1\}$: vanishing curvature and nonzero torsion.

One might ask why the different compatibility conditions, which apply to distinct spaces, have the same mathematical form. The following Section~\ref{sec:geo} will contain the full geometrical treatment with curvature and torsion. It will not be too difficult to see (mathematically) that the transition between the two spaces is provided by the expression of the spin connection~(\ref{2.30-1}). On the one hand, we can have the situation where the Levi-Civita connection vanishes, while on the other it is the spin connection that vanishes. This difference is captured by the frame fields and their first derivatives which in turn are related to our key geometrical quantities.

\section{Geometrical compatibility conditions}
\label{sec:geo}

\subsection{Geometrical identities}

The geometrical starting point for all compatibility conditions is the Bianchi identity which is satisfied by the curvature tensor and is given by
\begin{equation}
  \label{4.6}
  \nabla_\rho R^{ab}{}_{\mu\nu} +
  \nabla_\nu R^{ab}{}_{\rho\mu} +
  \nabla_\mu R^{ab}{}_{\nu\rho} =
  R^{ab}{}_{\tau\nu}T^\tau{}_{\mu\rho} +
  R^{ab}{}_{\tau\mu}T^\tau{}_{\rho\nu} +
  R^{ab}{}_{\tau\rho}T^\tau{}_{\nu\mu}\;,
\end{equation}
see for instance~\cite{JS1954}. For completeness, we also state the well-known identity
\begin{equation}
  \label{4.1}
  R^\rho{}_{\sigma\mu\nu} + R^\rho{}_{\mu\nu\sigma} + R^\rho{}_{\nu\sigma\mu} =
  \nabla_\sigma T^\rho{}_{\mu\nu} +
  \nabla_\mu T^\rho{}_{\nu\sigma} +
  \nabla_\nu T^\rho{}_{\sigma\mu} -
  T^\rho{}_{\sigma\lambda}T^\lambda{}_{\mu\nu} -
  T^\rho{}_{\mu\lambda}T^\lambda{}_{\nu\sigma} -
  T^\rho{}_{\nu\lambda}T^\lambda{}_{\sigma\mu}\;,
\end{equation}
for the Riemann curvature tensor which will also be required. Using $R^{ab}{}_{\mu\nu} = R^{\lambda\sigma}{}_{\mu\nu}e^a_\lambda e^b_\sigma$ and contracting twice over indices $\lambda$ and $\rho$, and $\sigma$ and $\nu$, gives the well-known twice contracted Bianchi identity
\begin{align}
  \nabla_\rho\left(R^\rho{}_{\mu}-\frac{1}{2} \delta^\rho_\mu R\right) =
  R^\lambda{}_{\tau} T^\tau{}_{\mu\lambda} +
  \frac{1}{2}R^{\lambda\sigma}{}_{\tau\mu} T^\tau{}_{\lambda\sigma}\;.
  \label{sec5:rr}
\end{align}
The term in the first bracket is the Einstein tensor so that the most general compatibility condition can be written as
\begin{equation}
  \label{4.7}
  \nabla_\rho G^\rho{}_{\mu} =
  R^\rho{}_{\tau}T^\tau{}_{\mu\rho} +
  \frac{1}{2}R^{\rho\sigma}{}_{\tau\mu}T^\tau{}_{\rho\sigma}\;.
\end{equation}
Equations~(\ref{4.6}) and~(\ref{4.1}) can be seen as a compatibility or integrability condition in the following sense. One cannot choose the curvature tensor and the torsion tensor fully independently as the above equations need to be satisfied for a consistent geometrical approach.

Let us now recall Eq.~(\ref{2.40}), the Einstein tensor in terms of $\Omega$, which was $G = \Curl\Omega + \Cof\Omega$. Next, we use the decomposition of the spin connection~(\ref{2.30-1}) into~(\ref{2.35}) to obtain
\begin{align}
  \Omega_{c\mu} = -\frac{1}{2}\omega_{\mu}{}^{ab}\varepsilon_{abc} =
  -\frac{1}{2}\left(\accentset{\circ}{\omega}_{\mu}{}^{ab} +
  K^{a}{}_{\mu}{}^b\right)\varepsilon_{abc} =
  \accentset{\circ}{\Omega}_{c\mu}+\Gamma_{c\mu}\;.
  \label{4.13}
\end{align}
When this decomposition is put into the explicit Einstein tensor equation, a slightly lengthy calculation yields
\begin{multline}
  G^{\lambda c} = (\Curl\Omega)^{c\lambda}+(\Cof\Omega)^{c\lambda}=
  \Curl(\accentset{\circ}{\Omega}+\Gamma)^{c\lambda}+
  \Cof(\accentset{\circ}{\Omega}+\Gamma)^{c\lambda}
  \\=
  (\Curl\accentset{\circ}{\Omega})^{c\lambda}+(\Curl\Gamma)^{c\lambda} +
  \frac{1}{2}\varepsilon^{cab}\varepsilon^{\lambda\mu\nu}
  \left(\accentset{\circ}{\Omega}_{a\mu}+\Gamma_{a\mu}\right)
  \left(\accentset{\circ}{\Omega}_{b\nu}+\Gamma_{b\nu}\right)
  \\=
  \Bigl\{(\Curl\accentset{\circ}{\Omega})^{c\lambda}+
  (\Cof\accentset{\circ}{\Omega})^{c\lambda}\Bigr\}+
  \Bigl\{(\Curl\Gamma)^{c\lambda}+
  (\Cof\Gamma)^{c\lambda}\Bigr\}+
  \varepsilon^{cab}\varepsilon^{\lambda\mu\nu}\accentset{\circ}{\Omega}_{a\mu}\Gamma_{b\nu}\;.
  \label{4.15}
\end{multline}
Let us note that the final term on the right-hand side can be written as
\begin{equation}
  \label{4.16}
  \accentset{\circ}{\Omega}_{a\mu} \Gamma_{b\nu} =
  -\frac{1}{2}\accentset{\circ}{\omega}_{\mu}{}^{p}{}_q\epsilon_{ap}{}^{q}\Gamma_{b\nu} =
  -\frac{1}{2}\epsilon_{ap}{}^{q}\left(e^p_\rho\accentset{\circ}{\Gamma}^\rho_{\mu\sigma}E^\sigma_q+e^p_\sigma\partial_\mu E^\sigma_q\right)\Gamma_{b\nu}\;,
\end{equation}
where we used relation~(\ref{2.9-1}) together with definition~(\ref{2.35}). We are now ready to present a complete description of compatibility conditions encountered so far following a unified approach using~(\ref{4.7}) and~(\ref{4.15}).

Before doing so, let us note the key property of the Einstein tensor decomposition~(\ref{4.15}). The final term is a cross-term which mixes the curvature and the torsion parts of the connection. Without this term one of the compatibility conditions would necessarily imply the other, it is precisely the presence of this term which gives the general condition a much richer structure.

\subsection{Compatibility conditions}

\subsubsection*{No curvature and no torsion}

Let us set $R^\rho{}_{\sigma\mu\nu}=0$ and $T^\lambda{}_{\mu\nu}=0$ in~(\ref{4.15}). Then we must also have $K_{b\nu}=\Gamma_{b\nu}=0$ by the definitions and we find the compatibility condition
\begin{equation}
  \label{4.8}
  \accentset{\circ}{G}^{\lambda c}=(\Curl\accentset{\circ}{\Omega})^{c\lambda}+
  (\Cof\accentset{\circ}{\Omega})^{c\lambda}=0\;,
\end{equation}
which is Vall\'ee's result~(\ref{3.6}) discussed earlier.

\subsubsection*{No curvature but torsion}

Let us set $R^\rho{}_{\sigma\mu\nu}=0$ and $T^\lambda{}_{\mu\nu} \neq 0$ in~(\ref{4.15}) which becomes
\begin{equation}
  G^{\lambda c}=\Bigl\{(\Curl\Gamma)^{c\lambda}+
  (\Cof\Gamma)^{c\lambda}\Bigr\}+
  \varepsilon^{cab}\varepsilon^{\lambda\mu\nu}\accentset{\circ}{\Omega}_{a\mu}\Gamma_{b\nu}=0\;.
\end{equation}
Furthermore if we impose the condition $U^a_\mu=\delta^a_\mu$ , then as observed in~(\ref{3.27-4}), $\accentset{\circ}{\Omega}_{a\mu}=0$ the compatibility condition reduces to
\begin{equation}
  \label{4.8a}
  G^{\lambda c} = (\Curl\Gamma)^{c\lambda} + (\Cof\Gamma)^{c\lambda} = 0\;.
\end{equation}
In this case we have Nye's result~(\ref{3.25}) which is equivalent to Skyrme's condition~(\ref{covcurl2}).

\subsubsection*{No torsion but curvature}

Using $R^\rho{}_{\sigma\mu\nu} \neq 0$ and $T^\lambda{}_{\mu\nu}=0$ in~(\ref{4.7}) we have the compatibility condition
\begin{equation}
  \label{4.12}
  \accentset{\circ}{\nabla}_\mu\accentset{\circ}{G}^{\mu\sigma}=0\;,
\end{equation}
where $\accentset{\circ}{G}^{\mu\sigma}$ is now a symmetric tensor. These equations are well known in the context of General Relativity (in this case one works on a four dimensional Lorentzian manifold) where they imply the energy-momentum conservation equations.

\subsubsection*{Curvature and torsion}

Let us now consider the general case where neither curvature nor torsion are assumed to vanish. In this case there are no `compatibility' equations as such to satisfy. However, one should read~(\ref{4.6}) and~(\ref{4.1}) as integrability or consistency conditions in the following sense: One cannot prescribe an arbitrary curvature tensor and an arbitrary torsion tensor at the same time, these tensor need to satisfy the relations~(\ref{4.6}) and~(\ref{4.1}), as already said in the above.

\subsection{An application to axisymmetric problems}

The compatibility conditions for an axisymmetric three dimensional continuum were reconsidered recently in~\cite{ML2020}. Using our geometrical approach shows, once more, the role played by geometrical objects in continuum mechanics. To study an axisymmetric material we choose the line element with cylindrical coordinate $X^\mu=\{r,\theta,z\}$ to be
\begin{align}
  ds^2 = (1+\epsilon_{rr}) dr^2 + r^2 (1+\epsilon_{\theta\theta})d\theta^2 +
  (1+\epsilon_{zz}) dz^2+ 2\epsilon_{rz} dr dz \,,
\end{align}
where the strain components $\epsilon_{\mu\nu}$ are functions of $r$ and $z$ only. Next, following the above, one now computes the Einstein tensor components $G_{\tau\lambda}$ while assuming $\epsilon_{\mu\nu} \ll 1$. It turns out that the incompatibility tensor $\mathbf{S}$ used in~\cite{ML2020} is identical to the three-dimensional Einstein tensor. This means we have
\begin{align}
  \accentset{\circ}{G}_{\tau\lambda} = \mathbf{S}_{\tau\lambda} =
  \bigl[ \nabla \times (\nabla \times \epsilon) \bigr]_{\tau\lambda} = 0\,.
  \label{einstein2}
\end{align}
The square brackets here indicate that we are referring to the components of the enclosed object. Furthermore, the Einstein tensor must satisfy identity~(\ref{4.12}) which means we find the neat relation
\begin{align}
  \accentset{\circ}{\nabla}^\tau\accentset{\circ}{G}_{\tau\lambda} = \bigl[ \nabla \cdot \mathbf{S} \bigr]_{\lambda} = 0\,.
\end{align}
The condition $\nabla\cdot\B{S}=0$ is valid for classical elasticity and does not necessarily apply to other more general settings. On the other hand, the identity $\accentset{\circ}{\nabla}^\tau\accentset{\circ}{G}_{\tau\lambda}=0$ crucially depends on the vanishing of the right-hand side of~(\ref{4.7}) and therefore on the specific model being considered. 

The equivalence of both results is expected as they follow from Bianchi type identities in geometry. It was then observed in~\cite{ML2020} that the four non-vanishing components of $\mathbf{S}$, or equivalently $G_{\tau\lambda}$, are not independent and that it should be possible to reduce this system further, this is then demonstrated. The three-dimensional Einstein tensor hence plays an important role in continuum mechanics. Further applications of the compatibility condition in solving nonlinear systems with non-trivial dislocations and disclinations in both classical and micropolar theories can be found in~\cite{ER1989, LZ1997, SD2011, AZ2018, EG2019}.

\section{Conclusions and discussions}

The starting point of this work was the use of geometrical tools for the study of compatibility conditions in elasticity. It is well known that the vanishing of the Riemann curvature tensor of the deformed body yields the compatibility conditions equivalent to the Saint-Venant compatibility conditions~\cite{BB1955, KK1963, SG1972, EK1980, HK1989, SG2003}, which are otherwise derived by considering higher order partial derivatives, that have to necessarily commute. Since the Riemann curvature tensor satisfies various geometrical identities it is expected that these identities also play a role in continuum mechanics. After revisiting these basic results, we were able to show that Vall\'{e}e's compatibility condition, which was also derived using tools of differential geometry, is in fact equivalent to the vanishing of the three-dimensional Einstein tensor. Our first key result was thus~(\ref{2.40}) which is also of interest in its own right as the representation of the Einstein tensor in this form appears to be new. The underlying geometrical space contained curvature and torsion which made it possible to apply our result to Nye's tensor and show the link to Skyrme's model, which is very well known in particle physics. Given that the determinant of the Nye tensor is related to a topological quantity, it is interesting to speculate about other links between topology and quantities used in continuum mechanics. A geometrical formulation, as much as is possible, will be key in understanding this.

As a small application, we applied our results to a recent study of the compatibility conditions for an axisymmetric problem, where we showed that the (linearised) Einstein tensor naturally appears and can be expressed as the double curl of the strain tensor~(\ref{einstein2}). This was our second representation of the Einstein tensor in an unusual way. It naturally led to additional identities that needed to be satisfied which then reduced the number of compatibility equations further.

Our study can be extended further by dropping our assumption of vanishing non-metricity and introducing the non-metricity tensor $Q_{\alpha\mu\nu} := \nabla_\alpha g_{\mu\nu}$. The polar decomposition of the tetrad will not be affected by this, however, the connection and spin connection components will change. For instance, the decomposition~(\ref{2.30}) will contain a third piece due to non-metricity which hence enter the Riemann curvature tensor. Its identities in turn will involve additional terms~\cite{JS1954} and it would be interesting to understand the compatibility conditions in this extended framework. In~\cite{AY2012-1} a geometry of this type, with non-vanishing non-metricity, was considered to study a distribution of point defects. The space in question was torsion free and did not contain curvature. It is not clear, at the moment, whether or not the Einstein tensor will play an important role in this setting as well and how non-metricity would affect the various conditions that were studied.\\

\section*{Acknowledgements}
We would like to thank Friedrich Hehl, Patrizio Neff and Jaakko Nissinen for helpful suggestions. Yongjo Lee is supported by EPSRC Doctoral Training Programme (EP/N509577/1).

%\bibliographystyle{unsrt}
%\bibliography{references}

\end{document}